# Variations of Interatomic Force Constants in the Topological Phonon Phase Transition of AlGaN

Daosheng Tang [1,2,*]

[1]School of Rail Transportation, Soochow University, Suzhou 215131, China

[2]School of Physical Science and Technology, Soochow University, Suzhou 215006, China

(October 16, 2023)

## Abstract

The topological effects of phonons have been extensively studied in various materials, particularly in the wide-bandgap semiconductor GaN, which has the potential to improve heat dissipation in power electronics due to its intrinsic, topologically-protected, non-dissipative phonon surface states. Nevertheless, the phase transition of the Weyl phonons in nitrides and their composite alloys has yet to be elucidated. To unveil the microscale origin, topological phonon properties in AlGaN alloys are investigated using the virtual crystal approximation (VCA) and special quasi-random structure (SQS) approaches in this work. It is found that phase transitions in Weyl phonons are evidently present in AlGaN alloys and nitride single crystals. Under strain states, both GaN and AlN show a more prominent phase transition of Weyl phonons when subjected to biaxial compressive and uniaxial tensile strains. And it has been observed that the *zz* components in the self-term and the transverse 1NN force constants (FCs) are the most influential during the phase transition. The nonlinear Weyl phonon transition in AlGaN alloys, as modeled by the VCA, is reflected in the normalized self-term and first-nearest-neighbor (1NN) FCs, which vary in a nonlinear fashion with an increasing magnitude. This nonlinear phenomenon is also confirmed in the SQS modeling, where the unfolded phonon dispersions are consistent with those in the VCA modeling. With increased branches, hundreds of Weyl phonons are present accompanied by significant disorders in normalized FCs, which mainly occur for N atoms in self-terms and for all components in normalized 1NN FCs.

[*]Corresponding author, Email: dstang@suda.edu.cn

The last few decades have seen a surge in the investigation of wide-bandgap semiconductors, such as SiC and III-nitrides (GaN and AlN), for their superior electronic properties, including high electron mobilities and breakdown voltages, *etc* [1-3]. The requirement for superior performance of wide bandgap semiconductor devices has resulted in a marked increase in working power and frequency, making heat dissipation a bottleneck issue. In the GaN high electron mobility transistor (HEMT), a very important GaN-based electronic, the near-junction thermal management becomes critical, where heat spreading and phonon ballistic-diffusive thermal transport are coupled and contribute the most to the total thermal resistance [4, 5]. In order to tackle the critical heat dissipation issues in GaN-based high-power electronics, first-principles calculations and thermal-physical experiments have been used to reveal the lattice thermal transport rules in single-crystal nitride semiconductors [6-21] as well as their alloy systems [10, 12, 22].

It has been recently reported that wurtzite III-nitride semiconductors contain topological phonons such as Weyl phonons and nodal ring phonons [8, 11, 23, 24], which are promising for realizing non-dissipative or low-dissipative phonon thermal transport at surfaces and interfaces and improving the near-junction thermal management significantly due to the topologically protected surface phonon states. However, different from the topological phonons protected by the crystalline and time reversal symmetry, these Weyl phonons are the results of accidental degeneracy and not protected by any specific lattice symmetry. Therefore, they may be affected by the disorders in atomic masses and force constants (FCs). Although the topological features of wurtzite structures have been identified as either trivial (*e.g.*, GaN and CuBr) or nontrivial (*e.g.*, AlN and CuI) [8, 24], the origin of the topological phonon phase transition in these structures is still not clear. Utilizing alloy systems with different element concentrations gives us a platform to uncover the mechanisms of phonon phase transitions. In this work, phonon dispersions and the topological phonon phase transition in AlGaN alloys are systematically calculated from first principles with the aid of both the Virtual Crystal Approximation (VCA) and Special Quasi-random Structure (SQS) methods. To further figure out the microscopic origin of these phenomena, the force constants, including the self-term and first-nearest-neighbor (1NN) term force constants, are particularly analyzed.

All the first-principles calculations are performed based on density functional theory as implemented in the open-source software Quantum ESPRESSO [25, 26] using projective augmented wave pseudopotential [27], along with the Perdew-Burke-Ernzerhof [28] implementation of the generalized gradient approximation for the exchange-correlation functional. The kinetic energy cutoff of the wave function is set as 50 Ry (680 eV). The lattice structures are optimized with the maximal Hellmann-Feynman force convergence threshold being $1\times10^{-9}$ Ry/bohr ($7.2\times10^{-9}$ eV/Å) on each atom. The Monkhorst-Pack scheme [29] is adopted to sample the first Brillouin zone with a $15\times15\times9$ mesh for nitrides single crystals of wurtzite structure. Both $k$-mesh sampling grids and kinetic energy cutoffs have been examined with careful convergence tests [7, 10, 11]. Second interatomic force constants (2nd IFCs) are extracted based on the finite displacement supercell approach as implemented in open-source software Phonopy [13], with a supercell size of $5\times5\times3$ for wurtzite structure and corresponding $3\times3\times3$ $k$-mesh grids. The Born effective charges and high-frequency dielectric constants are calculated by density functional perturbation theory for nonanalytical corrections of the dynamical matrix near the Gamma point in phonon dispersion calculations brought from polar effects [30]. Besides general phonon properties, the topological properties of phonon band structures are also calculated to unveil the topological phonon phase transition in wurtzite alloys. The Chern number of phonon bands is employed here to indicate the trivial or nontrivial topology of phonon crossing points, which is defined based on the phonon eigenvectors, and calculated by the Wilson loop method with the aid of open-source software WannierTools [31]. More calculation details can be found in Supplementary Materials and Refs [8, 11, 32, 33]. To model the alloy structures and their phonon properties, both the VCA method and SQS method [34] are employed in this work. In the VCA modeling, IFCs of corresponding single crystal including wurtzite AlN and GaN are calculated at first, which are used to perform VCA-based analyses, where the bond lengths, atomic masses of cation atoms, and IFCs of AlGaN alloys are determined based on the direct linear interpolations. The `mcsqs` algorithm [35] implemented in the Alloy Theoretic Automated Toolkit (ATAT) [36] is adopted to generate the 32-atom SQSs with different Al concentrations (See more details in Supplementary Materials). With these 32-atom supercells as new unitcells, finite displacement method is adopted to extract the 2nd IFCs with supercell sizes 3×3×2.

Phonon band structures of AlGaN alloys with varying Al concentrations can be obtained based on the lattice dynamics theory, as shown in Figure 1. Considering the significant computing requirements of SQS modeling with large supercells, three cases with Al concentrations of 0.25, 0.50, and 0.75 have been chosen. In Figures 1(a)-(c), phonon dispersions of AlGaN modeled by the VCA method varying with the Al concentrations are illustrated and compared with those in the SQS modeling. By comparing the phonon dispersions of AlGaN alloys modeled by the VCA and SQS methods, three significant differences can be found. The first one is that the acoustic branches is overestimated in the VCA modeling, while the optical branches with lower frequency (5-15 THz) are underestimated. Contrary to the predictions in highest frequencies in lower optical branches, the lowest frequencies in the higher optical branches (>15 THz) are nearly the same in the VCA and SQS modeling, which results in the overestimation of phonon bandgap in the VCA modeling. Since a large supercell is adopted to model the alloy in the SQS modeling, it results in a smaller first Brillouin zone. To make the comparison clearer, a phonon unfolding technique [37] is employed here to extract the unfolded phonon dispersions in the SQS modeling. By unfolding the phonon dispersions in large alloy supercell into the unitcell of single-crystals, it can be found that the unfolded phonon dispersions are much similar to the dispersions in the VCA modeling, in both phonon frequencies range and bandgaps as shown in Figures 1(d)-(f). The unfolded results also illustrate that phonon modes in Γ-M and Γ-A paths are still clear especially the acoustic branches while phonon modes in other paths have been dispersed significantly, indicating the disorder feature in the alloy system.

More information about phonon band structures is hidden in their geometry structures of eigenstates, known as the topological effects of phonons. The evolution of Weyl phonon properties in nitride single crystals with strain states and in AlGaN alloys with respect to the Al concentrations is illustrated by VCA and SQS modeling, as shown in Figure 2. In Figure 2(a), Weyl phonon points in the irreducible Brillouin zone (IBZ) of wurtzite GaN under different strain states are illustrated. The strain employed in this work is a 5% relative variation of lattice constants, with biaxial strain corresponding to the variation of the lattice constant *a* and uniaxial strain to *c*. It is noted here that though there is no topological feature of Weyl phonons in the Berry curvature for wurtzite GaN [8], the calculations of the Chern number show that there is a pair of Weyl phonon points in the acoustic branches located at the $k_z=0$ plane in the reciprocal space. Basically, the negative results in the Berry curvature imply the weakness of this pair of Weyl phonons. This work primarily focuses on the presence of Weyl phonons, rather than their characteristics, which is to be studied in a separate work. While the Weyl phonons in wurtzite nitrides result from an accidental degeneracy, the strain fields, including the biaxial and uniaxial strains, do indeed stimulate the phase transition of the phonons. In specific, under biaxial tensile and uniaxial compressive strain states, Weyl phonon points in GaN disappear, *i.e.*, GaN becomes a topologically trivial system. Under a biaxial compressive strain state, Weyl phonon points increase to three and the original pair moves closer to the boundary of the BZ. And the Weyl points in the phonon branches are still those in the acoustic branches. Additional phonon phase transitions occur under a uniaxial tensile strain state. Besides the movement of the original pair of Weyl phonons, a new pair of Weyl phonons in optical branches is present. By decreasing the lattice constant *a*, *i.e.*, under biaxial compressive or uniaxial tensile strain states, Weyl phonon points in wurtzite GaN will generally rise. Despite the changes in the number of Weyl phonons, their locations are still in the $k_z = 0$ plane in the reciprocal space.

The other nitride single crystal studied in this work, wurtzite AlN, shows different phonon phase transition behavior under strain states (Figure 2(b)). It has been confirmed in previous work that the wurtzite AlN is topologically nontrivial for phonons, with clear surface phonon arc states [8]. The Weyl phonon search calculations in the current work also show positive results, where six pairs of Weyl phonon points are present in the $k_z = 0$ plane in the IBZ. To be more precise, two pairs of Weyl points are related to acoustic phonon branches, and the other four pairs are associated with optical phonon branches. The application of biaxial or uniaxial strains leads to a decrease in the Weyl phonon points. By decreasing the lattice constant *c*, *i.e.*, under biaxial tensile and uniaxial compressive strain states, Weyl phonons at acoustic branches, as well as Weyl phonons at optical branches with small wave vectors, disappear. Under a uniaxial tensile strain state, a pair of Weyl phonons at the acoustic branch with a small wave vector and a pair of Weyl phonons at the optical branch with a large wave vector are still conserved. It's different from the previous scenario; under a biaxial compressive strain state, a pair of Weyl phonons at the acoustic branch and another pair of Weyl phonons at the optical branch with large wavevectors remain present. Moreover, both a biaxial compressive and tensile strain produce a pair of Weyl phonon points in the internal Brillouin zone located at neither high-symmetry paths nor planes, and this phenomenon is more significant in the biaxial compressive strain case.

Besides the topological phonon phase transition under strain states, Figure 2 also shows the transition due to the alloy in the framework of the VCA, from results in (a) to (e). Upon the rise in Al concentration, the original Weyl phonons in GaN persist and move nearer to the boundary of the Brillouin zone without disappearance or generation of new points. When the Al concentration increases to a large value, *e.g.*, 0.75, a new pair of Weyl phonons is produced in the optical phonon branches. With Al atoms occupying all the cation positions of the nitride single crystals, *i.e.*, AlN, an increased amount of Weyl phonons in both the acoustic and optical branches is observed. In fact, this phase transition modeled by the VCA method is an ideal process since the gradual changes in FCs and atomic masses cannot be realized under actual conditions. In Figures 2(c), (d), and (f), hundreds of Weyl phonons are detected in the full Brillouin zone of AlGaN alloys modeled by the SQS method. In SQS modeling, the primitive cell of the AlGaN alloy consists of 32 atoms with 96 phonon branches. As the chemical ratio of Al atoms increases in AlGaN alloys, the phonon frequencies will increase, yet the maximum frequency will remain lower than that of AlN. Within this finite frequency range, the increase of phonon branches significantly increases the number of phonon crossing points, many of which are topologically non-trivial and have non-zero Chern numbers. Due to the large supercell size and random distribution of Al atoms, no crystalline symmetry is present in the alloy system. In consequence, the IBZ is the same as the Brillouin zone. More detailed Weyl phonon information of AlGaN can be seen in the Supplementary Materials.

To further unveil the mechanisms for the differences in phonon properties of AlGaN alloys, as well as nitride single crystals, the variations of the FCs are analyzed. Figures 3-5 demonstrate different terms of FCs including self-terms and first-nearest-neighbor (1NN) terms. At strain states, bond lengths and interatomic FCs vary, yet atomic masses remain constant. For the self-terms in Figures 3(a) to (d), variations of FCs show an obvious anisotropy. The biaxial strain induces significant changes in the *xx* component while the largest two changes in the *zz* component result from the biaxial compressive strain and uniaxial tensile strain, corresponding to the case hosting the significant phase transitions. In GaN, the 1NN FCs mainly refer to the FCs between Ga and N, which are the first nearest neighbors to one another. For a Ga atom, there is one N atom directly above it and three N atoms around it in the plane perpendicular to the polar axis. For the sake of simplicity, the FCs in the former case are referred to as longitudinal FCs, and those in the latter case are referred to as transverse FCs. In Figures 4(a)-(c), 1NN FCs in GaN systems under different strain states are presented. Classified as 1NN atoms, four Ga-N pairs are present, three of them transverse and one longitudinal, and the bond length in the transverse pair is slightly shorter. It can be found from these data that the transverse FCs contribute the most to the total changes of FCs and are promising to affect much more in the topological phonon phase transition. Similar to the variations of FCs in GaN, the variations of the *zz* component of the self-term FCs in AlN are consistent with the phase transition in the phonon topology, where the effects of biaxial compressive and uniaxial tensile strains are more obvious (Figures 3(d)-(f)). And the transverse FCs of the 1NN atoms also show more significant changes (Figures 4(c) and (d)).

As the Al concentration increases from 0 to 1.00 in increments of 0.25, the atomic masses at the cation positions will vary accordingly. To render the FCs analyses more significant, the normalized FCs are employed; this is calculated by dividing the FCs by the squared mass products of the two concerned atoms. Figures 3(e) and (f) shows the variations of normalized self-term FCs in the VCA and SQS modeling, where the total number of atoms in the supercells for phonon calculations are 300 and 576, respectively. As the Al concentration increases, both the *xx* and *zz* components in the self-term normalized FCs for cations such as Al, Ga, and X (denoting the cations in the VCA modeling) atoms increase, with the magnitude of the variation also increasing. This approach is consistent with the phonon phase transition in AlGaN alloys modeled by the VCA method, where the Weyl phonon points first move closer to the boundary of the Brillouin zone and then new Weyl phonon points are generated. Figure 5 illustrates the results of normalized 1NN FCs; the variation magnitudes also increase with the increasing of the Al concentration, especially the *xx*, *yy*, and *zz* components indicated by the dotted lines, which is similar to that in the self-terms. Second-nearest-neighbor (2NN) FCs where corresponding atom pairs mainly refer to the X-X (Al-Al or Ga-Ga) and N-N atoms in the same primitive unitcell or the nearest unitcell, are also calculated in this work. Full data of normalized 1NN and 2NN FCs of AlGaN alloys in the VCA and SQS modeling can be found in the Supplementary Materials. Different from FCs in nitride single crystals and AlGaN alloys in the VCA modeling, FCs in AlGaN alloys in the SQS modeling show obvious disorder. Specifically, the disorder in self-term FCs is mainly observed in the N atoms, exhibiting a range of FCs from the magnitude of the FCs of the N atoms in GaN to that in AlN, while the self-term FCs of the Al atoms and the Ga atoms are much similar to those in the corresponding single crystals, with less disorder. For the normalized 1NN FCs of SQSs in Figures 5(d)-(f), both disorders in bond lengths and FCs are clearly illustrated, where both of them are distributed in a relatively wide range instead of single values in the VCA modeling. In particular, the data in three cases with different Al concentrations are severely overlapped. And the nonlinear variation trends shown in the VCA modeling are actually present in each configuration, as indicated by the green lines in the figures.

In summary, the phonon band structures and topological phonon phase transitions of AlGaN alloys are investigated from first principles using the VCA and SQS modeling. Topological phonon phase transitions, focusing on the Weyl phonon in strained GaN, AlN, and AlGaN alloys, are reported based on the Chern number and Wilson loop calculations. Biaxial compressive and uniaxial tensile strains have a pronounced effect on the phase transition of Weyl phonons in both GaN and AlN. It has been discovered that, during the phase transition of phonons under strain states in both GaN and AlN, the *zz* components in self-term and 1NN transverse force constants are the most influential. As Al concentration increases, the Weyl phonons in AlGaN alloys, as modeled by the VCA, move closer to the boundary of the Brillouin zone, and a new pair of Weyl phonons in optical branches is created when the concentration is large. Consistent with the nonlinear Weyl phonon transition in AlGaN alloys, the normalized self-term and 1NN FCs also vary in a nonlinear form, with an increasing magnitude of change that is also confirmed by the SQS modeling. In the more realistic modeling by the SQS method, a large supercell introduces more phonon branches accompanied by more topologically nontrivial crossing points, *i.e.*, Weyl phonons. The significant variations and disorders in normalized FCs mainly occur for N atoms in self-terms and for all components in normalized 1NN FCs.

**Supplementary Materials**

See the supplementary materials for additional details of the VCA and SQS modeling, the calculation of the Chern number, and for additional data.

**Acknowledgement**

This work was financially supported by the National Natural Science Foundation of China (No. 52206105), the China Postdoctoral Science Foundation (No. 2021M702384), Jiangsu Funding Program for Excellent Postdoctoral Talent (No. 2022ZB594), and Natural Science Foundation of the Jiangsu Higher Education Institutions of China (No. 22KJB470008).

**Conflict of Interest**

The authors declare no conflict of interest.

**Data Availability Statement**

The data that support the findings of this study are available from the corresponding author on reasonable request.

**Figure captions**

**Figure 1**. (a)-(c) Phonon band structures of AlGaN alloys by the VCA and SQS modeling, (d)-(f) Unfolded phonon band structures of AlGaN alloys by the SQS modeling.

**Figure 2**. Weyl phonon phase transition in $Al_xGa_{1-x}N$ alloys modeled by the VCA and SQS methods (a) $x=0.0$ (GaN), (b) $x=1.00$ (AlN) (c) $x=0.25$, (d) $x=0.50$, and (e) $x=0.75$. The black and blue lines denote the Brillouin zone and irreducible Brillouin zone, respectively. Red circular and blue triangular dots represent the Weyl points at acoustic and optical branches, respectively.

**Figure 3**. Diagonal terms of force constants in strained GaN (a) *xx*, (b) *zz*, in strained AlN (c) *xx*, (d) *z,* and in $Al_xGa_{1-x}N$ alloys (e) *xx*, (f) *zz*. In (e) and (f), solid points indicate the normalized FCs in the VCA modeling, and hollow points signify the normalized FCs in the SQS modeling. Different Al concentrations are identified by the colors of the cases.

**Figure 4**. 1NN force constant distribution in (a)-(c) strained GaN and (d)-(f) strained AlN. The lines in (a) and (b) are used to emphasize the transverse bonding.

**Figure 5**. Normalized 1NN force constant distribution in $Al_xGa_{1-x}N$ alloys by the (a)-(c) VCA and (d)-(f) SQS modeling. The lines in figures are used to illustrate the nonlinear variations of FCs.

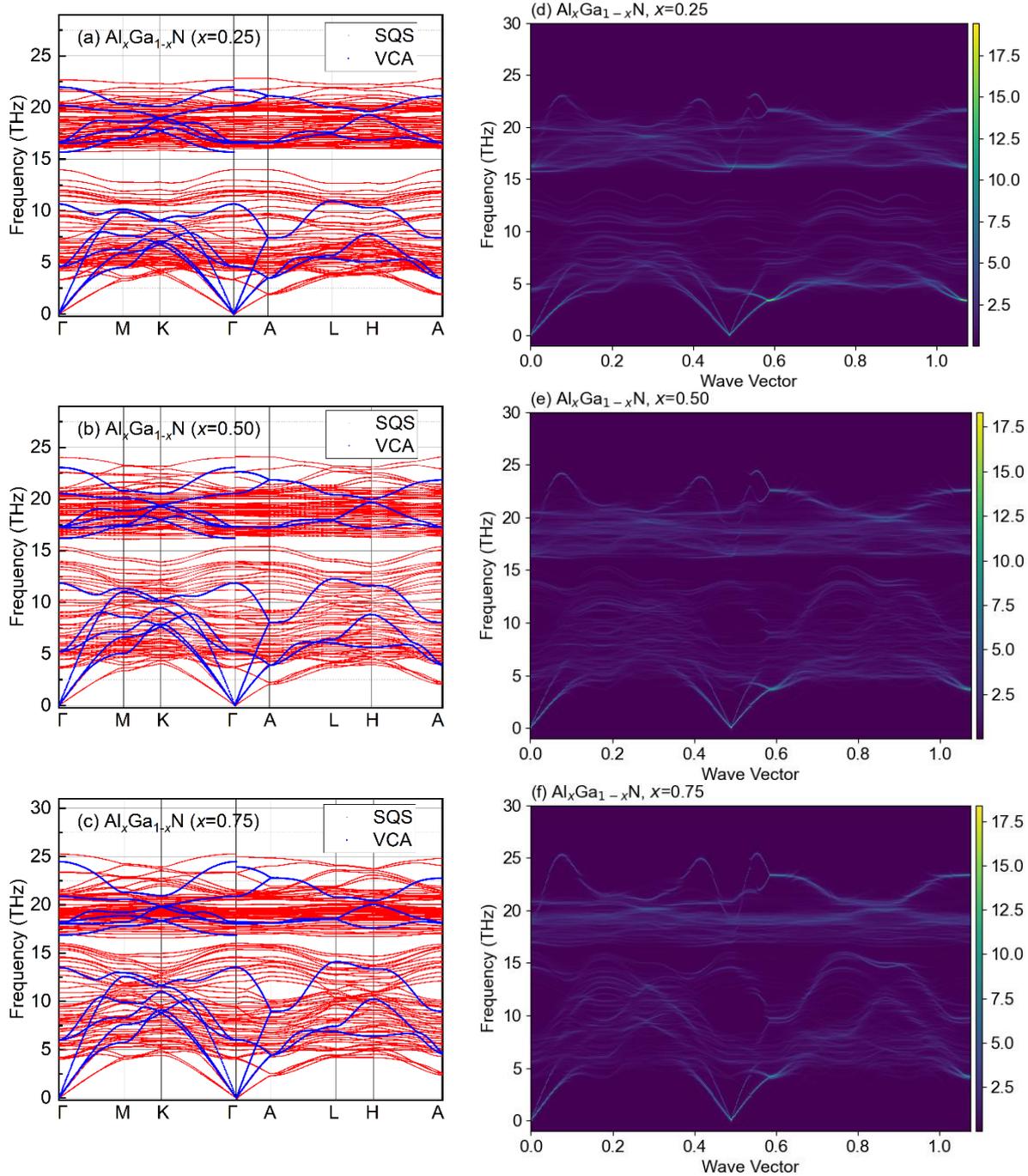

Figure 1. (a)-(c) Phonon band structures of AlGaN alloys by the VCA and SQS modeling, (d)-(f) Unfolded phonon band structures of AlGaN alloys by the SQS modeling.

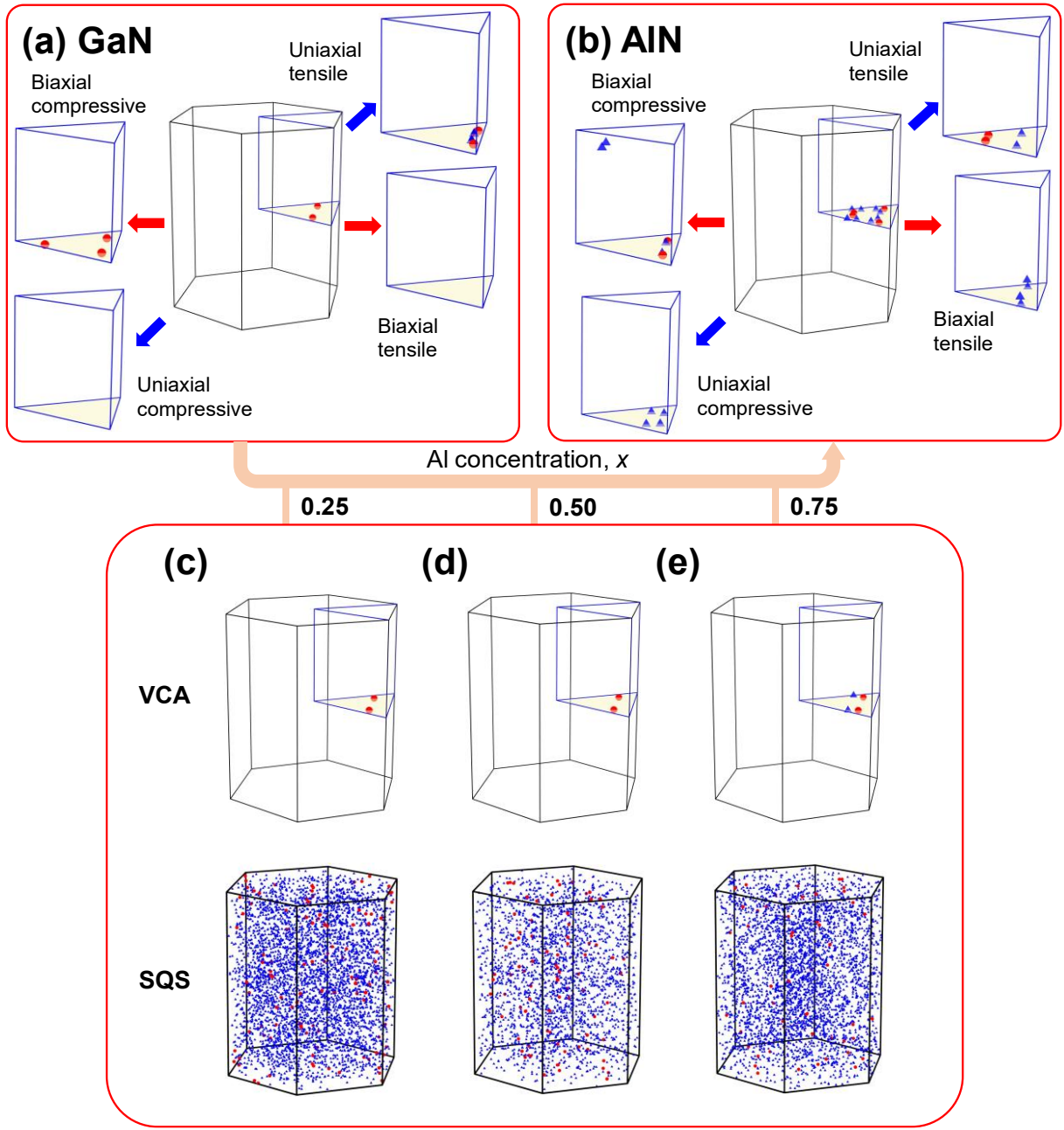

Figure 2. Weyl phonon phase transition in Al$_x$Ga$_{1-x}$N alloys modeled by the VCA and SQS methods (a) $x$=0.0 (GaN), (b) $x$=1.00 (AlN) (c) $x$=0.25, (d) $x$=0.50, and (e) $x$=0.75. The black and blue lines denote the Brillouin zone and irreducible Brillouin zone, respectively. Red circular and blue triangular dots represent the Weyl points at acoustic and optical branches, respectively.

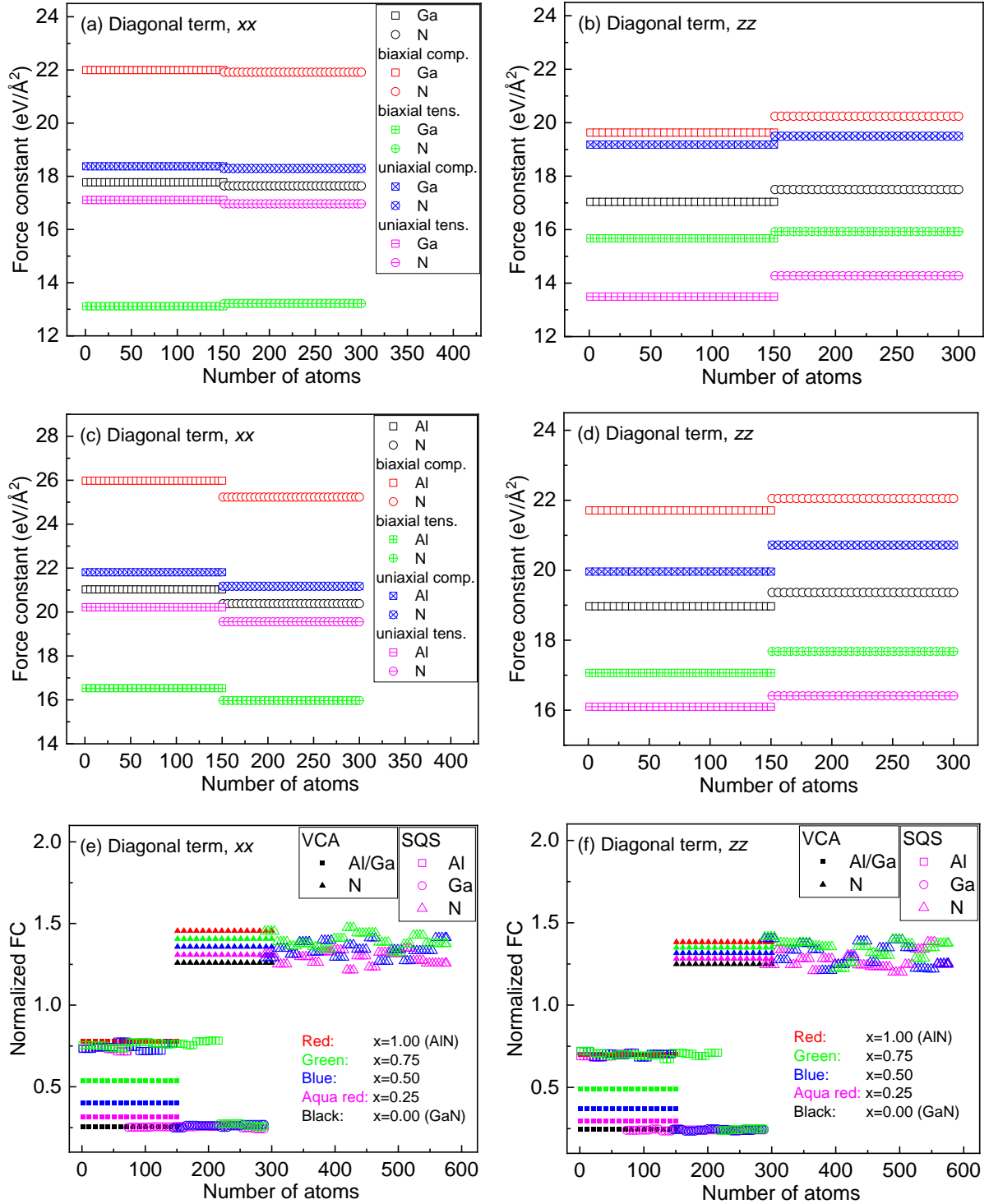

Figure 3. Diagonal terms of force constants in strained GaN (a) *xx*, (b) *zz*, in strained AlN (c) *xx*, (d) *z*, and in Al$_x$Ga$_{1-x}$N alloys (e) *xx*, (f) *zz*. In (e) and (f), solid points indicate the normalized FCs in the VCA modeling, and hollow points signify the normalized FCs in the SQS modeling. Different Al concentrations are identified by the colors of the cases.

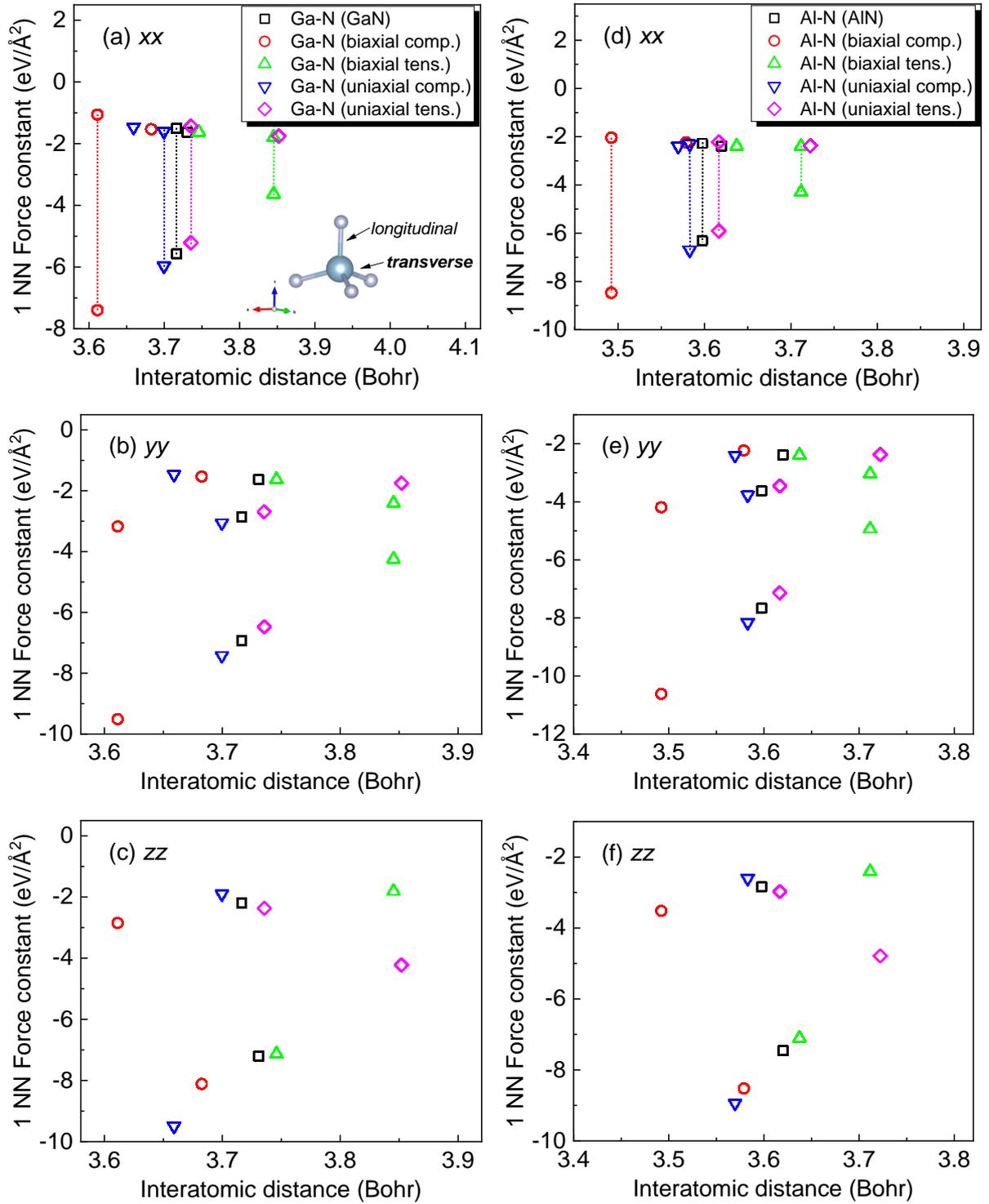

Figure 4. 1NN force constant distribution in (a)-(c) strained GaN and (d)-(f) strained AlN. The lines in (a) and (b) are used to emphasize the transverse bonding.

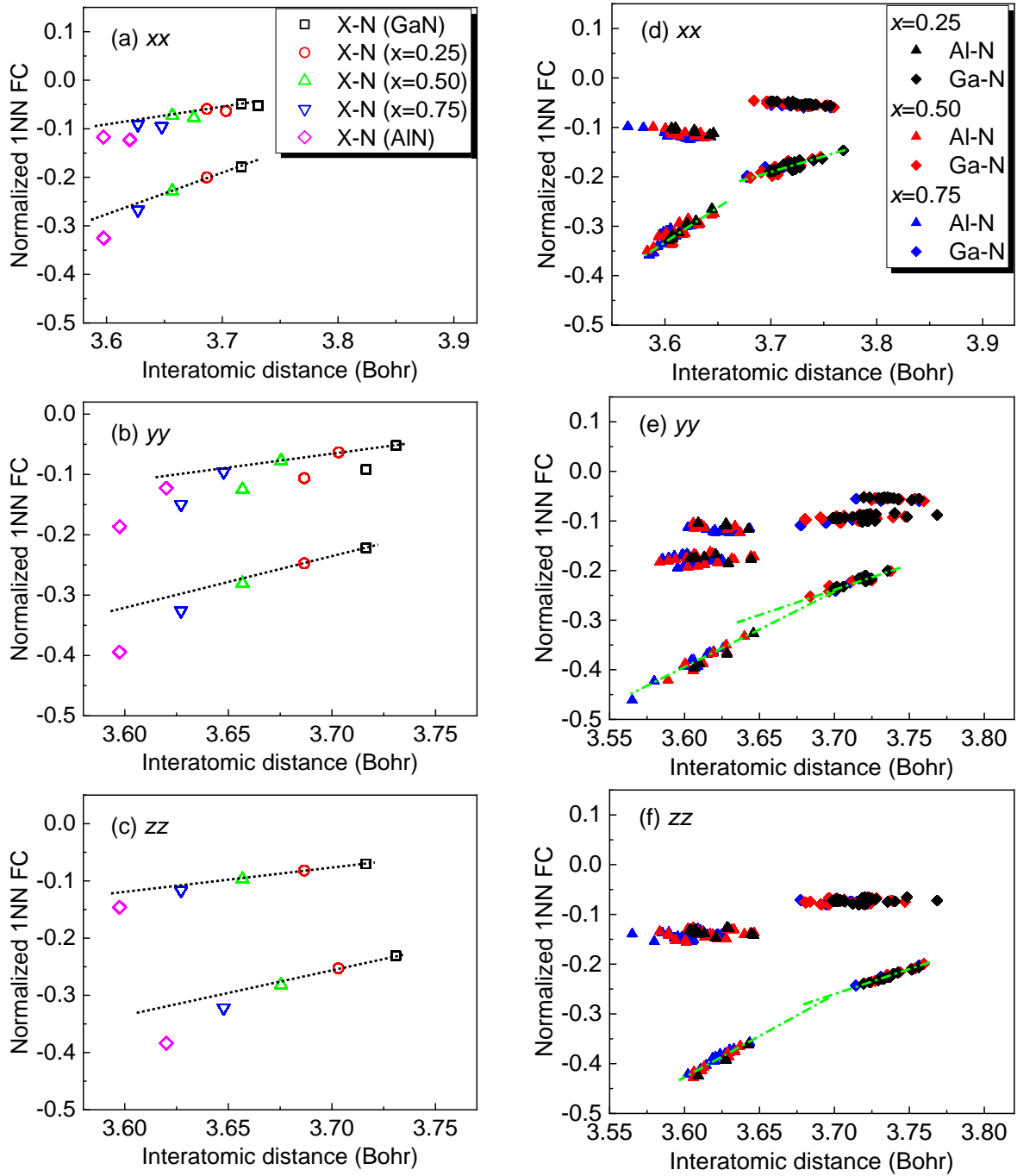

Figure 5. Normalized 1NN force constant distribution in Al$_x$Ga$_{1-x}$N alloys by the (a)-(c) VCA and (d)-(f) SQS modeling. The lines in figures are used to illustrate the nonlinear variations of FCs.